\documentclass[12pt]{article}
\usepackage{amssymb}
\usepackage{fontenc}
\usepackage{times}
\usepackage{mathptmx}
\usepackage{graphicx}

\begin{document}
\begin{titlepage}
\noindent
DESY 13-196 \hfill Nov 2013\\
LPN 13-071 \\

\begin{center}
  {\bf 
    \Large 
    Top-quark Pair Production in a Running Mass Scheme
  }
  \vspace{1.5cm}

  {\large
    M. Dowling$^{\,a}$ 
    and
    S. Moch$^{\,a,b}$
  }\\
  \vspace{1.2cm}

  {\it 
    $^a$Deutsches Elektronensynchrotron DESY \\
    Platanenallee 6, D--15738 Zeuthen, Germany \\
    \vspace{0.2cm}
    $^b$ II. Institut f\"ur Theoretische Physik, Universit\"at Hamburg \\
    Luruper Chaussee 149, D--22761 Hamburg, Germany \\
  }
  \vspace{2.4cm}



\large
{\bf Abstract}
\vspace{-0.2cm}
\end{center}
Between the Tevatron and LHC, top-quark physics is now becoming an area for precision physics. This has lead to an increase in theoretical activity to match the experimental accuracy of top anti-top production. We discuss the difficulty in properly defining the top-quark mass as measured by experiments and present results for differential distributions of top-quark pair production in a running mass scheme. The use of such a scheme shows better convergence in the perturbative expansion and improves the scale dependence as opposed to the typical on-shell scheme.
\vfill
\end{titlepage}
\section{Introduction}
The top-quark provides a unique window into Quantum Chromodynamics (QCD) due to its large mass and very short lifetime.
It's lifetime is so short that it decays well before hadronizing and thus experiments have access to properties that would not normally be measurable for individual quarks.
In addition, the very short decay time means that the production and decay of top quarks can be treated almost completely in perturbation theory to higher orders.
The top sector then becomes an area for precision tests of QCD.

One of the properties that has been measured very precisely is the mass of the top quark.
The most recent combined measurements from the Tevatron and LHC are
\begin{eqnarray}
\textrm{Tevatron \cite{CDF:2013jga}:} \quad m_t & =  &173.2 \pm 0.87 \textrm{GeV}, \\
\textrm{LHC \cite{ATLAS:2012coa}:} \quad m_t & = &173.3 \pm 1.4 \textrm{GeV}. 
\end{eqnarray}
There are of course many other measurements using various techniques and decay channels but these give a feel of how precisely the experiments are able to determine the mass.

One common element of these two measurements is that they are so-called "direct" measurements.
This effectively means that the mass is obtained by comparing various event properties with Monte Carlo (MC) simulations of what should be seen.
By fitting the simulations to the data, a value of the mass can be extracted.
There are various techniques for actually doing this: including the Template Method \cite{Abe:1994st}, the Matrix Element Method \cite{Abazov:2004cs}, and the use of Ideograms \cite{Abazov:2007rk}.
These methods are similar in that they depend on matching the data to a MC simulation of what should be seen for various values of the top mass.
One important consideration when doing these mass measurements is: Which mass is being measured?

\section{Top-quark Mass}
In the Standard Model (SM), mass is a free parameter. The bare mass of a particle that appears in the Lagrangian is an infinite quantity which needs to be adjusted by an infinite renormalization contribution to give a physical value. i.e. the value measured in the lab.
The choice of the renormalization scheme affects the value of the mass obtained, usually, in a well defined way.
Typically when considering a particle's mass it is the pole mass that is being referred to.
This is effectively the mass of the particle that would be measured if it was free and corresponds to the location of the pole of the propagator.
It is this definition of mass that is usually assumed to be measured by experiments because it is the mass used in the perturbative calculations that act as input.

There are a few problems with this though.
For one thing, quarks are not free particles and thus it does not make much sense to talk about a pole mass for a quark.
The second problem is that because the measurements rely on MC simulations, there is some dependence on the models that are used.
Examples of the model dependent effects that need to be included are colour recombination and bound state effects.
For this reason, it was proposed that what is really being measured is a MC mass which is related to the pole mass perturbatively \cite{Hoang:2008xm} via 
\begin{equation}
m_{pole} = m_{MC} + Q_0[\alpha_S(Q_0)c_1 + ...].
\end{equation}
It is then typically argued that the scale $Q_0$ should be about 1GeV which is the scale of the cutoff in radiation in parton shower evolution.
In this case $\alpha_S$ is on the order of 1 and $c_1$ is completely unknown.
To obtain an estimate of the error introduced by this relation, it is assumed that $c_1$ should also be on the order of 1 giving a total uncertainty of about 1GeV.
This is on the same order as the current combined measurement from the Tevatron and hence it is important to further study this relation.
In addition to the unknown definition of the MC mass, perturbation theory in the pole mass scheme applied to the top-quark suffers from the infrared renormalon.
This limits the accuracy to $\mathcal{O}(\Lambda_{QCD})$ \cite{Bigi:1994em}.

Finally the top quark is not a stable particle which means that the on-shell calculations currently used are missing finite width effects.
Flagari {\it et al.}, \cite{Falgari:2013gwa}, studied the off-shell effects in the differential production cross-section with respect to the invariant reconstructed top mass to show that the contributions can have a significant effect on the determination of the top mass.
\begin{figure}[ht]
\centering
\includegraphics[width=0.6\textwidth]{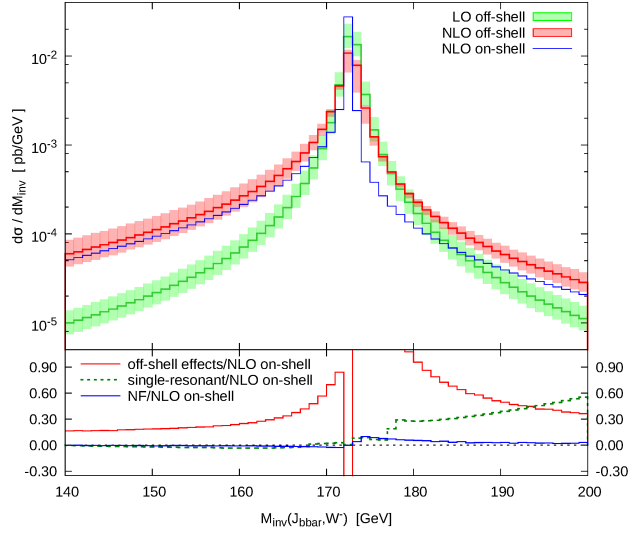}
\caption{\label{fig:off shell} The differential cross-section with respect to the invariant mass of a reconstructed top. In blue is the NLO on-shell differential cross-section and in red includes the off-shell effects.}
\end{figure}
Figure \ref{fig:off shell} shows that taking into account the off-shell-ness of the top quark can possibly lead to sizeable effects that need to be taken into account when determining the mass. More detailed studies into these effects are ongoing.

\section{Methods of Reducing Uncertainty}
As a way of checking the direct top mass measurements, a number of other observables are being considered.
These observables either reduce the dependence or are completely independent of the various uncertainties discussed in the previous section.
Recently, CMS used the endpoints of kinematic distributions to extract a mass for the top quark \cite{Chatrchyan:2013boa}.
This method relies on theoretical descriptions of the endpoints of various kinematic distributions to simultaneously extract values for the neutrino, W-boson and top quark masses.
Using the world average values for the neutrino and W-boson masses, 0 GeV and 80.4 GeV respectively, they find
\begin{equation}
m_{pole} = 173.9 \pm 0.9 \textrm{(stat.)}^{+1.7}_{-2.1}\textrm{(syst.) GeV}
\end{equation} 
in agreement with other top pole mass determinations.

\begin{figure}[ht!]
\centering
\includegraphics[width=0.8\textwidth]{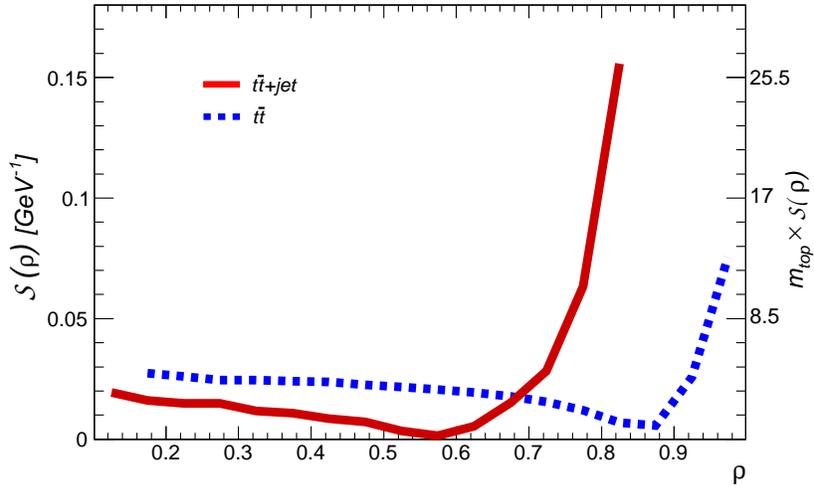}
\caption{\label{fig:sensitivity} Estimates of the sensitivity to the top-quark mass in the $t\overline{t}$ (blue) and $t\overline{t}+1jet$ (red) systems.}
\end{figure}

Another option that has been proposed is to use the differential distribution for the production of a $t\overline{t}$ pair plus one jet \cite{Alioli:2013mxa}.
The NLO corrections to this process are known meaning that the mass of the top-quark is well defined.
In addition, it was argued that this observable could be competetive in precision when extracting a top-quark mass.
In \cite{Alioli:2013mxa} the authors were able to show that an approximate relation between the error in the measurement and the error in the pole mass is given by
\begin{equation}
\left| \frac{\Delta \mathcal{R}}{\mathcal{R}} \right| \approx
      \bigg( m_{pole}\times S(\rho) \bigg)
      \left| \frac{\Delta m_{pole}}{m_{pole}}\right|,
\end{equation}
where $\mathcal{R}$ is defined by,
\begin{equation}
\mathcal{R}(m_{pole},\rho) = \frac{1}{\sigma_{t\overline{t}+1jet}} \frac{d\sigma_{t\overline{t}+1jet}}{d\rho}(m_{pole},\rho),
\end{equation}
$\rho = \frac{2m_0}{\sqrt{s_{t\overline{t}j}}}$, $S(\rho)$ is the sensitivity and $m_0$ sets the scale of the top mass.
Figure \ref{fig:sensitivity} shows the sensitivity to the pole mass as a function of $\rho$ for the inclusive $t\overline{t}$ differential cross-section and the $t\overline{t}+1jet$ differential cross-section.
It is seen that using the 1-jet description increases the sensitivity in the region just below threshold making it accessible at least from a theory standpoint.
It should be noted that these curves are only estimates and further studies of the sensitivity are underway.

The final option that will be discussed is to use the measured production cross-section to obtain the mass.
This benefits from the fact that the NNLO corrections to the production cross-section are known \cite{Czakon:2013goa} and the cross-section can be measured in experiments in an unambiguous way (for example with a counting experiment).
With current accuracy though, the error in the extracted top-mass is currently much larger than those using the more conventional direct measurements.
This can be seen in the analysis of Tevatron data, where the $\overline{\textrm{MS}}$ mass was extracted and translated to the pole mass for comparison \cite{Alekhin:2012py}:
\begin{eqnarray}
m_{\overline{MS}} & = & 163.3\pm 2.7\textrm{GeV}, \\
m_{pole} & = & 173.3\pm 2.8\textrm{GeV}.
\end{eqnarray}
As can be seen, the pole mass agrees well with the values obtained using direct methods but with a slightly larger error.

\section{The $\overline{\textrm{MS}}$ Scheme}
\begin{figure}[ht]
\centering
\includegraphics[width=0.49\textwidth]{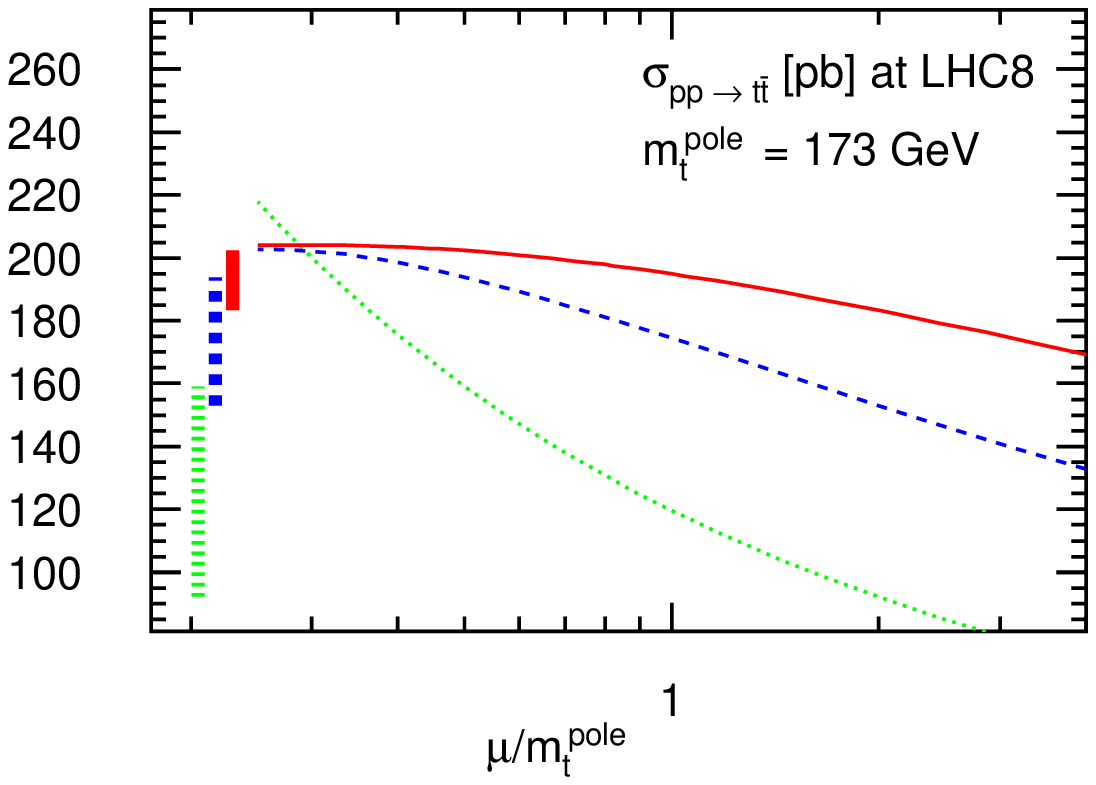}\,
\includegraphics[width=0.49\textwidth]{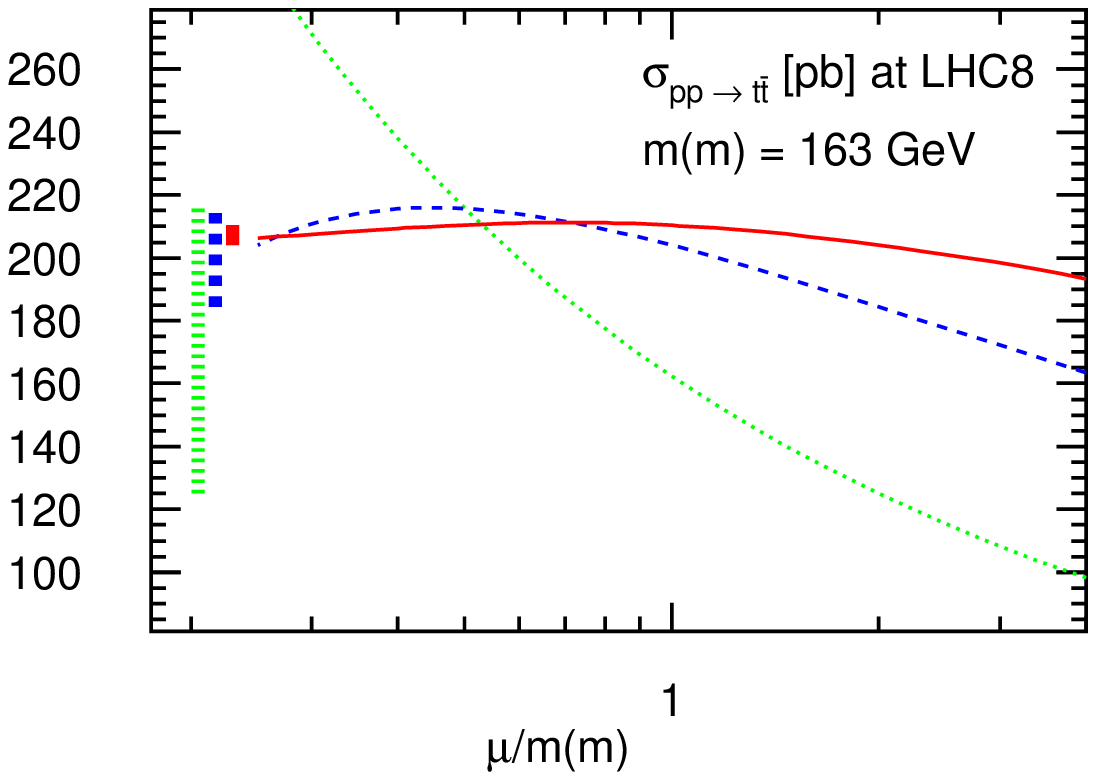}
\caption{\label{fig:series}The scale dependence of the $t\overline{t}$ production cross-section in the pole mass scheme (left) and $\overline{\textrm{MS}}$ scheme (right) at LO, NLO, and NNLO. The vertical bars indicate the variation in the range $\mu/m_{pole}\in [1/2,2]$.}
\end{figure}

The authors of \cite{Alekhin:2012py} chose to use the $\overline{\textrm{MS}}$ mass instead of the pole mass for a few reasons.
The first is that the renormalon ambiguity mentioned earlier is not present in the $\overline{\textrm{MS}}$ scheme.
A second and perhaps more important reason comes from looking at the perturbative series describing the production cross-section.
Figure \ref{fig:series} shows the LO, NLO and NNLO scale dependence in the pole mass scheme as compared to the $\overline{\textrm{MS}}$ scheme.
We see that at NLO and NNLO, the variation of the scale dependence becomes significantly smaller. 
At NNLO, the variation in the pole mass scheme in the standard range $\mu/m_{pole} \in [1/2,2]$ amounts to $\Delta\sigma_{\textrm{NNLO}}=^{+3.8\%}_{-6.0\%}$.
In the $\overline{\textrm{MS}}$ this reduces to $\Delta\sigma_{\textrm{NNLO}}=^{+0.1\%}_{-3.0\%}$.
In addition to scale dependence, the perturbative series also shows an improvement in convergence.
Including the NNLO corrections in the pole mass scheme represents approximately a 12\% increase in the cross-section.
Comparing this to the $\overline{\textrm{MS}}$ scheme, it is seen that the NNLO corrections represent only a 3\% increase in the cross-section.

In addition to the total cross-section, these improvements hold for differential cross-sections.
For convenience, we define the diffrential-cross section with respect to $X$ as
\begin{equation}
\label{eq:diff}
\frac{d\sigma}{dX} = \left(\frac{\alpha_S}{\pi}\right)^2\frac{d\sigma^{(0)}}{dX} + \left(\frac{\alpha_S}{\pi}\right)^3\frac{d\sigma^{(1)}}{dX} + \mathcal{O}(\alpha_S^4)
\end{equation}
With the number of top-quarks being produced at the LHC, experiments are starting to measure the differential cross-sections so a study of the improvements obtained by moving to the $\overline{\textrm{MS}}$ scheme is required.
In \cite{Dowling:2013baa} we have computed the differential cross-sections in the $\overline{\textrm{MS}}$ for transverse momentum, $p^t_T$, rapidity, $y^t$, and the invariant mass of the $t\overline{t}$ system, $m^{t\overline{t}}$.
The translation from the on-shell calculations to the $\overline{\textrm{MS}}$ scheme is obtained using the perturbative relation between the two schemes,
\begin{equation}
m_{pole} = \overline{m}(\overline{m})\left(1 + \frac{\alpha_s}{\pi}d_1 + \left(\frac{\alpha_s}{\pi}\right)^2d_2 + \mathcal{O}(\alpha_s^3)\right).
\end{equation}
When applied to the description of the differential cross-section, it is found that
 \begin{eqnarray}
      \frac{d\sigma(\overline{m}(\overline{m}))}{dX} &=& 
      \left(\frac{\alpha_s}{\pi}\right)^2\frac{d\sigma^{(0)}(\overline{m}(\overline{m}))}{dX}
      + \left(\frac{\alpha_s}{\pi}\right)^3 \left\{ 
      \frac{d\sigma^{(1)}(\overline{m}(\overline{m}))}{dX} \right. \nonumber \\
      & &
      \left.+ d_1 \overline{m}(\overline{m})
      \frac{d}{dm_t}\left(\frac{d\sigma^{(0)}(m_t)}{dX}\right)\biggr|_{m_t=\overline{m}(\overline{m})}
      \right\} + {\cal O}(\alpha_s^4),
    \end{eqnarray}
where $X$ is the variable of interest.
The extra derivative term, as compared to Equation (\ref{eq:diff}), causes a reduction in the contribution from the $\alpha^3$ term ultimately leading to the increased convergence in the perturbative series.
The required derivative terms have been computed analytically where possible.
In the cases of the $p^t_T$ and $y^t$ spectra, a partial derivative of the PDF contributions was required and carried out numerically.

\begin{figure}[ht]
\centering
\includegraphics[width=0.48\textwidth]{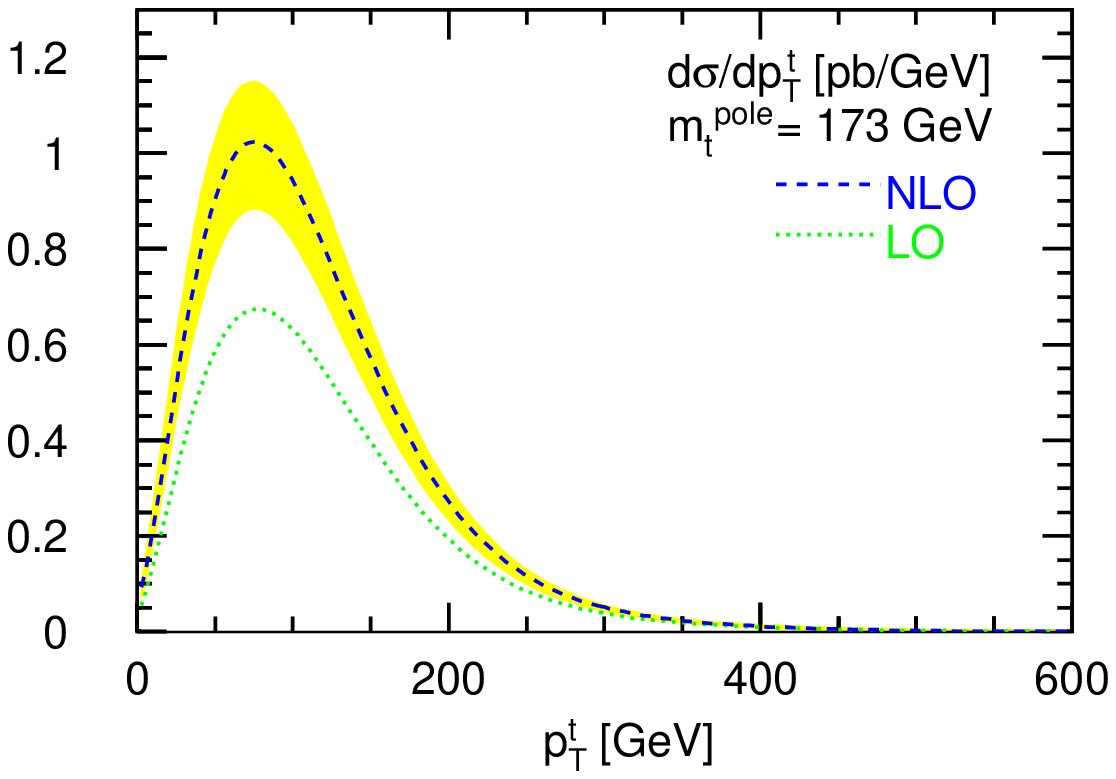}
\includegraphics[width=0.48\textwidth]{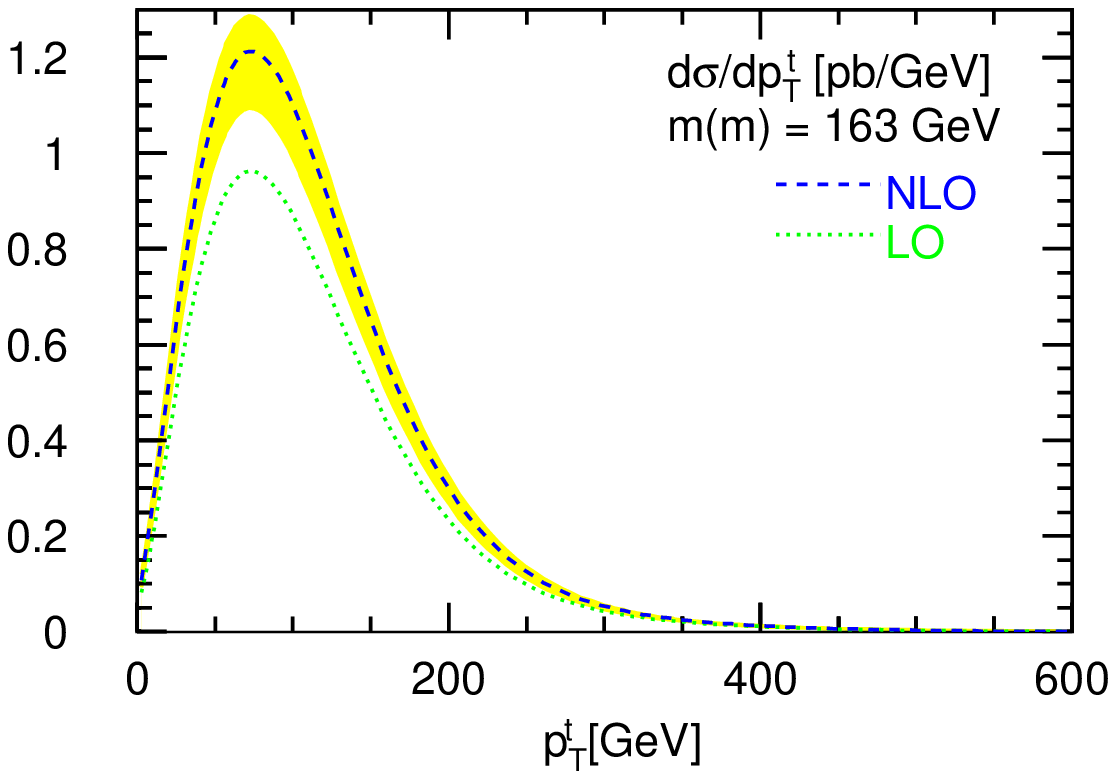}
\caption{\label{fig:pt}The $p^t_T$ differential-cross sections in the pole mass scheme (left) and $\overline{\textrm{MS}}$ scheme (right) at $\sqrt{S}=7 \textrm{TeV}$.}
\end{figure}

As an example of the results, consider the $p^t_T$ spectra shown in Figure \ref{fig:pt}.
We see that, as expected the difference between the LO (green) and NLO (blue) differential cross-sections, as well as the scale dependence(yellow) are smaller.
In addition, there is an overall shift of events towards the threshold region which results in a more pronounced peak.
At a value of $p_T = 75\textrm{GeV}$, the ratio $\sigma_{NLO}/\sigma_{LO}$ for the $p^t_T$ spectrum goes from 1.52 in the pole mass scheme to 1.26 in the $\overline{\textrm{MS}}$ scheme.
A detailed discussion of the results presented in this section can be found in \cite{Dowling:2013baa}.

\section{Conclusions}
The problem of defining a pole mass for the top-quark has been discussed, as well as the difficulties in associating the experimentally measured masses with a pole mass.
In order to deal with these problems, the relation between the MC and pole mass is being studied.
At the same time, other observables are being considered that are able to obtain independent measurements of the top-mass.
These other observables are able to circumvent many of the obstacles found in properly defining the top-mass being measured but, so far, do not provide the same level of precision found in direct measurements.
Finally, it has been shown how using the $\overline{\textrm{MS}}$ scheme improves the convergence of the perturbative series as well as the scale dependence in the theoretical predictions.
This may help to improve the measurements of the top-quark mass.

There is still a lot of work to be done in this area.
Other effects need to be included such as higher order corrections in the differential cross-sections, finite width effects and color-reconnection.
In particular, it has recently been found that electroweak corrections to the on-shell-$\overline{\textrm{MS}}$ relation largely cancel with the corresponding QCD corrections \cite{Jegerlehner:2012kn} for a Higgs boson with mass $m_H \sim 125\textrm{GeV}$.
It is however currently unclear as to how these corrections will affect the differential distributions presented here.

At a potential future $e^+e^-$ collider, the theory side is under slightly better control as $\textrm{(N)}^3\textrm{LO}$ corrections have been approximated  (see for example \cite{Hoang:2003ns}) and it has been suggested that it will be possible to determine the top-quark mass with a precision of about 100MeV.
\subsection*{Acknowledgments}
This work is partially supported by the Deutsche Forschungsgemeinschaft in Sonderforschungs\-be\-reich/Transregio~9
and by the European Commission through contract PITN-GA-2010-264564 ({\it LHCPhenoNet}).

\end{document}